\documentstyle[12pt]{article}
\def\lpmb#1{\mbox{\boldmath $#1$}}
\def\bLam{{\lpmb {\Lambda}}}
\def\{{\lbrace}
\def\}{\rbrace}
\def\<{\langle}
\def\>{\rangle}
\def\ie{{\it i.e.,}}
\def\eg{{\it e.g.,}}
\def\etal{{\it et al.}}

\def\lam{\lambda}
\def\bp{{\bf p}}
\def\bP{{\bf P}}
\def\bx{{\bf x}}
\def\bd{{\bf d}}
\def\H{{\cal H}}
\def\N{{\cal N}}
\def\D{{\cal D}}
\def\S{{\cal S}}
\def\Poincare{Poincar\'e }
\def\half{{\textstyle{1\over2}}}
\def\thalf{{\textstyle{3\over2}}}
\def\fourth{{\textstyle{1\over4}}}
\def\one{{\rm 1}}

\def\beq{\begin{equation}}
\def\eeq{\end{equation}}
\def\beqarray{\begin{eqnarray}}
\def\eeqarray{\end{eqnarray}}
\def\doeack{This work was supported by the Department of Energy, Nuclear
      Physics Division, under
      contracts W-31-109-ENG-38 (F.C.) and DE-FG02-86ER40286  (W.P.)}
\newcommand{\sect}[1]{\addtocounter{section}{1}\noindent
        \large{\bf\arabic{section}.~ #1}\vspace*{4mm}}

\date{April 1993}
\begin{document}
\title{{\small{\rightline{ANL-PHY-7524-TH-93}}}\vspace{10mm}
Vacuum Structures in Hamiltonian Light-Front Dynamics}
\author{F. Coester$^1$  and W. Polyzou$^2$}
\maketitle
\centerline{$^1${\it Physics Division, Argonne National Laboratory, Argonne IL
60439}}

\centerline{$^2${\it Department of Physics and Astronomy,
 The University of Iowa,
 Iowa City, IA 52242}}
\vspace{1cm}
\begin{abstract}
Hamiltonian light-front dynamics of quantum fields may provide
a useful approach to systematic non-perturbative approximations
to quantum field theories. We investigate inequivalent
Hilbert-space representations of the light-front field algebra in which
the stability group of the light-front  is implemented by unitary
transformations.
The Hilbert space representation of states  is generated by the operator
algebra
from the vacuum state. There is a large class of vacuum states besides the Fock
vacuum which meet all the invariance requirements. The light-front Hamiltonian
must annihilate the vacuum and have a positive spectrum. We exhibit relations
of the Hamiltonian to the nontrivial vacuum structure.
\end{abstract}

\newpage

\sect{Background}
\normalsize

In the usual formulation of quantum mechanics of finite systems kinematics and
dynamics are cleanly separated.
An operator algebra of coordinate, momentum and spin operators
describes the degrees of freedom of the system. The  Hilbert-space
representation of this algebra is unique within unitary
equivalence.\cite{Bogol}
The dynamics is specified by the Hamiltonian operator which is a function
of the dynamical variables. Many different dynamical laws exist for any
particular Hilbert space of states and the associated operator algebra.
In particular states of interacting and noninteracting particles
are in the same Hilbert space and the operator algebra  is not altered
by the dynamics.
This picture changes qualitatively for infinite systems. There are inequivalent
Hilbert-space representations of the canonical commutation relations.
\cite{VanH} - \cite{Segal} The Fock-space representation is only one of
infinitely
many Hilbert-space representations, with different vacuum  structure. The
vacuum
structure depends on the dynamics. The essential ingredients  are  vacuum
polarization
terms in the Hamiltonian and invariance under translations.
The Lorentz invariance of quantum field theories is not relevant in this
context.\cite{Coester,Araki}

In light-front Hamiltonian dynamics\cite{Dirac} there are important
differences from the usual equal-time picture.
The kinematic symmetries are the
transformations which leave the null plane $x^+:=t+x_3=0$ invariant.
The generator, $P^+$, of translations tangential to the light cone has a
positive
continuous spectrum $[0,\infty]$ and a point eigenvalue $0$ for the vacuum
state.
The most important consequence is the suppression of
vacuum polarization terms in the Hamiltonian.\cite{Weinberg}
This feature indicates that a Fock-space representation of quantum
field theories may provide a framework for systematic nonperturbative
approximations. Discretizations and Tamm-Dancoff truncations of the Fock space
have been
investigated for that purpose.\cite{Pauli}-\cite{Tang} In discretized
formulations the
operator $P^+$ has a discrete spectrum. In the expansion of fields modes with
$P^+=0$,
(\ie ``zero-modes'') can play an important role. The dynamics  generates
constraint equations which serve to eliminate these modes.\cite{Maskawa} -
\cite{Pinsky}

The standard approach consists of quantization of  classical field theories.
We believe it is preferable to consider quantum theory as the fundamental
theory
and construct quantum field theory at the outset as translationally
invariant quantum theory.
\cite{Rohrlich,Coester.NP}
Relativistic quantum mechanics of infinitely many free particles is easily
formulated
in a Hilbert space which is the infinite direct sum of tensor products of
one-particle
Hilbert spaces. (Fock space) Field operators  constructed
from particle creation and annihilation
operators are precisely the usual free fields and the generators of the
\Poincare
group obtained in this manner are identical to those obtained by integrating
appropriate
components of the energy-momentum tensor over the null plane $x^+=0$.
The Fock vacuum is
the only state vector in this Hilbert space which is invariant under
translations. The free field Hamiltonian annihilates the vacuum.
If a  Hamiltonian with interactions also annihilates the vacuum we have the
same pattern
as in quantum mechanics. The Hilbert space of states
and the operator algebra are the same for systems with and without interaction.

{}From the analogy of field algebras with Euclidean kinematics
\cite{Araki,Araki.W}
we expect that there are inequivalent Hilbert space representations of the
null-plane field algebra
required by classes of Hamiltonians with different properties.
In this context it is essential to observe that the construction of field
operators
described above  does  not yield operator valued functions  of points on the
light front but linear functionals (distributions) over suitable test functions
defined on the null plane $x^+=0$.\cite{Schlieder}-\cite{Gelfand} Important
qualitative differences from the Euclidean case are associated with
the stability group of the null plane and the distinctive
features of the appropriate test functions.

Interactions are introduced formally by adding to the free-field energy
momentum tensor
an interaction  term which can be obtained from a local  Lagrangean  density.
The energy-momentum tensor so
constructed is an operator valued functional of functions over the
four-dimensional
space-time, but it is not an operator functional of functions restricted to the
null
plane.
Thus the formal Hamiltonians obtained from local Lagrangeans  are not
operators
unless they are ``regularized''.  The cut-offs involved in the regularization
always
destroy  the \Poincare covariance, but the null-plane symmetry can be
preserved in many ways. Different regularizations may imply  different
vacuum structures.

The central question is whether there are classes
of regularized Hamiltonians which can provide systematic nonperturbative
approximations
to the observable consequences of the full field theory.
Here we will be concerned only with  preliminary questions  concerning the
existence of inequivalent
Hilbert space representations of the null-plane field algebra and the
associated
vacuum structures. It  is  vital to regularize the Hamiltonians of the local
field theories in a manner which preserves the nontrivial structure of the
physical vacuum.

The purpose of this paper is to exhibit inequivalent Hilbert space
representations
of the free-field operator algebra and the associated  vacuum structures.
Invariance and the cluster properties are essential features of these vacuum
structures,
each of which supports  different  classes of Hamiltonians.
We do not address the questions related to limiting procedures designed to
restore
\Poincare invariance.  Our  approach in this paper is exploratory.
While the language of
algebraic field theory\cite{Haag} is essential for the exposition of the
relevant ideas
we do not attempt a mathematically rigorous treatment at this point.
The basic assumptions  that specify the null-plane field algebra are formulated
in Section 2. This is essentially an adaptation of the well-known Wightman
axioms \cite{Streater,Haag.II}  and of the basic assumptions
of algebraic quantum field theory.\cite{Haag.III} The vacuum structures are
discussed in Section 3.\\

\vspace{5mm}
\sect{Null-Plane Fields}
\setcounter{equation}{0}
\normalsize

For any point $x$ in Minkowski space we use the light-front components
$x^+:=x^0+x^3$, $x^-:=x^0-x^3$ and $x_\perp=\{0,x_1,x_2,0\}$.
 For points in the null plane $x^+=0$,
denoted by $\bx:=\{x^-,x_\perp\}$, the scalar product $p\cdot x$ reduces to
\beq
p\cdot x =p_\perp\cdot x_\perp - \half p^+x^-\; .
\eeq
The null-plane component $p^+$  is positive if $p$ is a timelike
four-momentum with positive energy. The null plane $x^+=0$ is invariant
under a subgroup $\N$ of the \Poincare group which consists of the translations
$\bx\to \bx+\bd$ and the Lorentz transformations $\bLam(a,b)$ with the
$SL(2,C)$ representation
\beq
\bLam(a,b) \to \pmatrix{a&0\cr b& 1/a \cr}\; ,
\label{AB}
\eeq
and the space-time inversion $\bx \to -\bx$. We use the notation
$\{\bd,\bLam\}$ for
the transformation $\bx \to \bLam(\bx+\bd)$, and for any function $f(\bx)$
\beq
\{\bd,\bLam\}f(\bx) := f[\bLam(\bx+\bd)]\; .
\eeq
The group $\N$ is the kinematic subgroup of null-plane dynamics.
When the Euclidean group in three dimensions is the kinematic group
every hyperplane $x^0=$const. is invariant.
It is a distinctive  feature of null-plane dynamics that only the null plane
$x^+=0$ is invariant under $\N$. The null planes $x^+=$ const.$\not=0$ are
invariant only under a subgroup of $\N$.

\vspace{5mm}
\noindent {\bf 2.1 Polynomial Algebra  of Field Operators}
\vspace{3mm}

The properties of null-plane fields can be described by a series of assumptions
which follow the pattern of the Wightman axioms

\begin{itemize}
\item [A.] {\it Assumptions of Null-Plane Quantum Theory}

The states of the theory are described by unit rays in a separable Hilbert
space
$\H$. The kinematic transformations are given by a continuous unitary
representation of the stability group $\N$ of the null plane
$\{\bd,\bLam\} \to U(\bd,\bLam)$.
Since $U(\bd,\one)$ is unitary it can be written as
$U(\bd,\one)=\exp(i\bP\cdot\bd)$, where $\bP:=\{P^+,P_\perp\}$ is an unbounded
self-adjoint operator interpreted as the light-front momentum operator of the
theory. The spectrum of $P^+$ is  positive, $P^+\geq 0$.
({\it spectral condition})
The infinitesimal generators of the kinematic Lorentz transformations, $\bLam$,
are  $K_z$ and $\vec E$ for the light-front boosts and $J_z$ for the rotations
about the
longitudinal axis.

There is an invariant state, $\Omega$: $U(\bd,\bLam)\Omega =\Omega$,
unique up  to a constant phase factor.

\item [B.] {\it Assumptions about the Domain  of the Field Operators}
\begin{enumerate}
\item Fields are ``operator valued distributions'' over the null plane,
\beq
\phi(f):=\int d^3 \bx \phi(\bx)f(\bx)=\int d^3 \bp \phi(-\bp)f(\bp)
\eeq
where $d^3\bx:=\half dx^-d^2x_\perp$  and $d^3\bp:= dp^+d^2p_\perp$.
The test functions $f(\bx)$ are real valued functions related to the functions
 $f(\bp)=\bar f(-\bp)$  by the Fourier transform
\beqarray
&&f(\bx) = (2\pi)^{-\thalf}\int d^3\bp f(\bp)e^{i\bp\cdot\bx}\cr\cr
&&f(\bp) = (2\pi)^{-\thalf}\int d^3\bx f(\bx)e^{-i\bp\cdot\bx} \; .
\eeqarray
A suitable test function space $\hat \S$ has been defined by
Schlieder and Seiler.\cite{Schlieder}
By their definition  $f(\bx)\in \hat {\cal S}(\Re^3)$ are
derivatives of functions in the space $\S$ defined by Laurent Schwartz.
\cite{Schwartz,Gelfand}
The functions $f(\bp)\in \hat \S$ are  Schwartz functions
in $\S(\Re^3)$ multiplied by $p^+$.

\item The operators $\phi(f)$ and their
adjoints $\phi^*(f)$ are  unbounded operators
acting on $\H$ defined on some dense domain $\D\subset \H$.
The domain $\D$ is a linear set containing the vacuum state $\Omega$,
$ \Omega \in \D$.
The ranges of operators $U(\bd,\bLam)$, $\phi(f)$ and $\phi(f)^*$
are  in $\D$,
\beq
U(\bd,\bLam) \D \subset \D\; , \qquad \phi(f) \D \subset \D\; ,\qquad
\phi^*(f) \D \subset \D\; .
\eeq
If $\Psi_1, \Psi_2 \in \D$, then $(\Psi_1,\phi(f)\Psi_2)$ is a
linear functional of $f$, which satisfies
\beq
\<\Psi_1\vert\phi^*(f)\Psi_2\>= \overline{\<\Psi_2\vert \phi(f)\Psi_1\>}\; .
\eeq
\item In general there will be several types of fields
(distinguished by an index $i$) each with
tensor or spinor components (labeled by an index $\lambda$). We have then
\beq
\phi(f):=\sum_i\int d^3 \bx \phi^i_\lam(\bx)f^{i\lam}(\bx)\; .
\eeq
\end{enumerate}
%\newpage
\item[C.]{\it Transformation Law of the Fields}

The fields transform according to
\beq
U^{-1}(\bd,\bLam)\phi^i_\lam(\bx)U(\bd,\bLam)=
S_\lam^{(i)\rho}(\bLam)\phi^i_\rho(\bLam^{-1}\bx-\bd) \; ,
\eeq
where $S_\lam^{(i)\rho}(\bLam)$ is a finite dimensional representation
matrix of the Lorentz transformation $\bLam$.

\item[D.]{\it Canonical Commutation relations, Local Commutativity}

Charged scalar fields  satisfy the  canonical commutation relations,
\beq
[\phi(f),\phi^*(g)]=(f,g)-(g,f)\; ,\qquad [\phi(f),\phi(g)]=0 \; ,
\eeq
where the scalar product of the test functions is defined by
\beq
(f,g):= \int {d^3\bp\theta(p^+)\over p^+}\bar f(\bp)g(\bp)\; .
\eeq

If  $\phi(f)$ is a spinor field we have the corresponding anticommutation
relations
\beq
[\phi(f),\phi^\dagger(g)]_+=(f,g)+(g,f)\; ,\qquad [\phi(f),\phi(g)]_+=0
\eeq
where  the invariant inner product of the spinor test functions is
\beq
(f,g):=\int {d^3\bp\theta(p^+)\over p^+} f^\dagger(\bp) g(\bp)\; .
\eeq
Here $\phi^\dagger(g):=\phi^*(g)\beta$ and $f^\dagger(\bp):=\bar f(\bp)\beta$
and $\beta\equiv \gamma^0$ is the usual Dirac matrix.\\

For Hermitean scalar fields,$\phi^*(f)=\phi(f)$,  we have,
\beq
[\phi(f),\phi(g)]=\half[(f,g)-(g,f)]\; .
\label{CCR}
\eeq
or
\beq
[\phi(\bp),\phi(\bp')]={\delta(\bp+\bp')\over 2p^+}\; .
\label{CCRP}
\eeq

Two points, $\bx$ and $\bx'$, on the null plane are relatively spacelike
if \hfill\break $\vert x_\perp-x'_\perp \vert >0$.
The scalar product $(f,g)$ vanishes  when
the supports of the test functions $f$ and $g$ are
relatively spacelike.

\item[E.] {\it Completeness}

Polynomials of the operators $\phi(f)$ operating on the vacuum state
$\Omega$ generate a dense set of states in $\H$.

\item[F.] {\it Dynamics}

There exists a self-adjoint Hamiltonian operator $H\equiv P^-$ whose domain
contains the domain
$\D$ of Assumption B. The Hamiltonian
operator $P^-$ is invariant under the null-plane translations, $[P^-,\bP]=0$.
The operator $P^-$ together with the generators,$\bP$, of the null-plane
translations
transforms as a four-vector under the null-plane Lorentz transformations
(\ref{AB}). The mass operator, $M^2:=P^+P^-  - P_\perp^2$ is invariant and has
a
positive spectrum.
The spectrum of the Hamiltonian is the positive real axis.
The Hamiltonian operator annihilates the vacuum.
\end{itemize}

The following   discussions will be restricted throughout to Hermitean scalar
fields
satisfying the commutation relations (\ref{CCR}). We assume that the operator
algebra
is not modified by the dynamics.
With the definitions
\beq
\phi^{(\pm)}(f):= \int d^3\bp\phi^{(\pm)}(\bp)f(-\bp)\;, \qquad
\phi^{(\pm)}(\bp):=\theta(\pm p^+)\phi(\bp)
\; ,
\eeq
it follows from the commutation relations (\ref{CCRP}) and the definition
of the test function space that
\beq
[\phi^{(+)}(f),\phi^{(+)}(g)]=[\phi^{(-)}(f),\phi^{(-)}(g)]=0\; ,
\eeq
and
\beq
[\phi^{(+)}(f),\phi^{(-)}(g)]= \half (f,g)\; .
\label{CCPM}
\eeq
We can therefore define the ``normal ordered product'' of field operators
in the usual manner, \eg
\beqarray
:\phi(f_1)\phi(f_2):\;\; &:=&\phi^{(+)}(f_1) \phi^{(+)}(f_2)
+\phi^{(-)}(f_1) \phi^{(-)}(f_2)\cr\cr &+&\phi^{(-)}(f_1) \phi^{(+)}(f_2)
+\phi^{(-)}(f_2)\phi^{(+)}(f_1) \; .
\eeqarray

{}From  the kinematic covariance of the field and the spectral condition
it follows that
\beq
[P^+,\phi(f)]\Omega =\int d^3 \bp p^+\theta(p^+) \phi(-\bp) f(\bp) \Omega\; ,
\eeq
and hence
\beq
[P^+,\phi^{(+)}(f)]\Omega= \int d^3 \bp p^+\theta(p^+) \phi(\bp) f(-\bp) \Omega
=0\; ,\qquad
{\partial \phi^{(+)}(\bx)\over\partial x^-}\, \Omega =0\; .
\label{PPL}
\eeq
The operators $\phi^{(+)}(f)$
annihilate the vacuum for all test functions $f(\bp)$ whose  support excludes
some
neighborhood of $p^+=0$. If we assume that  $\phi^{(+)}(f)$ annihilates the
vacuum
for all $f$  then the vacuum is the Fock vacuum and we have the usual
Fock-space representation of
the field algebra.
Any nontrivial vacuum structure must be associated
with vanishing values of  $p^+$.

The generators of the light-front symmetries are obtained  from integrals
over the light-front components of the energy momentum tensor,
\beqarray
&&P^+= \int d^3 \bx\, T^{++}(\bx)\; , \qquad P_\perp= \int d^3 \bx\,
T^{+\perp}(\bx)\cr\cr
&&K_3= \int d^3 \bx\, T^{++}(\bx)x^-\; ,\qquad \vec E =\int d^3 \bx
\,T^{++}(\bx)x_\perp\cr\cr
&&J_3=\int d^3 \bx \,[T^{+2}(\bx)\,x_1-T^{+1}(\bx)\,x_2]\; ,
\eeqarray
where
\beq
T^{++}(\bx)= 4:{\partial \phi(\bx)\over \partial x^-}
{\partial \phi(\bx)\over \partial x^-}: \; ,
\eeq
and
\beq
T^{+\perp}(\bx)= 2 :{\partial \phi(\bx)\over \partial x_\perp}
{\partial \phi(\bx)\over \partial x^-}:\; .
\eeq
The kinematic generators annihilate vacuum even when there is a nontrivial
vacuum structure
at $p^+=0$.

According to Assumption F we have
\beq
P^-= {M^2+P^2_\perp\over P^+}\; ,
\eeq
and in analogy to (\ref{PPL})
\beq
[P^-,\phi^{(+)}(f)]\Omega =P^-\phi^{(+)}(f)\Omega = 0 \; .
\label{PPM}
\eeq
For the free-field  Hamiltonian,
\beq
P^- = \int d^3\bx T^{+-}(\bx)= \int d^3\bx
:\phi(\bx)[m^2-\nabla_\perp^2]\phi(\bx):
\eeq
we get from eq.~(\ref{PPM})
\beq
[P^-,\phi^{(+)}(f)]\Omega
= \int d^3 \bp {m^2+p_\perp^2\over p^+}\theta(p^+) \phi(\bp) f(-\bp) \Omega
=0\; .
\label{PMF}
\eeq
It follows that  the free-field vacuum is the Fock vacuum, which by definition
is
annihilated by $\phi^{(+)}(f)$ for all $f$. In general the condition
(\ref{PPM}) may
specify annihilation operators which require a zero-mode structure of the
vacuum.
A similar characterization of the vacuum structure by the annihilation
operators
was found in the Euclidean case \cite{Coester}.

The most convenient description of general vacuum structures results from
the formulation of the field algebra as an abstract $^*$-algebra\cite{Haag.III}
described in the next subsection.

\vspace{5mm}
\noindent{\bf 2.2 Abstract $^*$-Algebra }
\vspace{3mm}

The properties of field operators listed above can be derived from
the abstract $^*$-algebra  generated by
finite linear combinations,
\beq
R:=\sum_n c_n U(f_n)
\eeq
with complex coefficients $c_n$
and  Schlieder-Seiler test functions $f_n(\bx)$.
Here the unitaries $U(f)$  have the algebraic
properties and the norm topology of linear combinations of the Hilbert-space
operators
$\exp [i\phi(f)]$  considered in the previous subsection.
We have thus a
$C^*$ algebra,
\beq
\Vert R^*R\Vert = \Vert R \Vert^2\; ,
\eeq
with the following properties:
\begin{itemize}
\item[I.]{\it Involution}
\beq
R^*=\left[\sum_n c_n U(f_n)\right]^*=\sum_n \bar c_n U(-f_n)\; .
\eeq
\item[II.]{\it Product and Unit Element}
\beq
U(g)U(f)=U(f+g)e^{\fourth [(f,g)-(g,f)]}\; , \qquad  U(0)= \one \; .
\label{PROD}
\eeq
\end{itemize}
With the notation $U(f_1,f_2):=U(f)$, where $f(\bp)=f_1(\bp)-if_2(\bp)$ and
$f_(\bp), f_2(\bp)$ are real, we see the the algebra specified by
eq.~(\ref{PROD}) is the Weyl  algebra, \cite{LKS}
\beq
U(g_1,g_2)U(f_1,f_2)=U(g_1+f_1,g_2+f_2)\,e^{{i\over 2}[(f_2,g_1)-(f_1,g_2)]}\;
{}.
\eeq
\begin{itemize}
\item[III.]{\it Null-Plane Symmetry}

The  continuous null-plane symmetries are implemented by *-automorphisms of the
abstract algebra,
\beq
\{\bd,\bLam\}U(f) \to U(\{\bd,\bLam\}f)\; ,
\eeq
where
\beq
\{\bd,\bLam\}f(\bx)= f[\bLam(\bx+\bd)]\; .
\eeq
The space-time inversion $\bx \to -\bx$ is implemented  by the
anti-automorphism
\beq
cU(f) \to \bar c U(-\tilde f)\; ,
\eeq
where $\tilde f(\bx):= f(-\bx)$.

\item[IV.]{\it The Vacuum Functional}

The vacuum state is given by a continuous linear functional over the
algebra, $E[R]=\sum_ic_iE[U(f_i)]$, $E(f):=E[U(f)]$, with the properties:
\begin{enumerate}

\item  Positivity: $E[R^* R]=\sum c_n \bar c_m E(f_n-f_m)
e^{\fourth[(f_n,f_m)-(f_m,f_n)]} \geq 0$.

\item  Normalization: $E(0)=E[1] =1$.

\item  Reality: $\bar E(f)=E(-f)$

\item Invariance: $E(f)=E(\{\bd,\bLam\}f)\; , \quad E(f)=E(\tilde f)$

\item  Cluster Property:
$\lim_{\lambda\to \infty}[E(f+\{\lambda \bd\} g)-E(f)E(g)]=0$, $d_\perp\not=
0$.

\end{enumerate}
\end{itemize}

For a vacuum functional with these properties
the Gelfand-Naimark-Segal (GNS) construction \cite{Haag.III} provides a
Hilbert-space representation of the algebra:
\beq
\<\Omega\vert R\Omega\> =E[R]\; ,
\eeq
and for any two state vectors $\Psi_1:= R_1\Omega$, $\Psi_2:= R_2\Omega$
\beq
\<\Psi_2\vert \Psi_1\> = E[R_2^*R_1]\; .
\eeq
There is only one translationally invariant vector, $ \Omega$, in the Hilbert
space so constructed. (See ref. \cite{Araki} theorem 6.1)

\vspace{5mm}
\sect{Vacuum Structure}
\setcounter{equation}{0}
\normalsize

{}From the strong continuity of the operators $U( f)$ follows
 the existence of
self-adjoint field operators $\phi(f)$ such that
$U(f)=e^{i\phi(f)}$.
The explicit form of the Fock vacuum functional,
\beq
E_0(f):= e^{-\fourth (f,f)}
\eeq
follows immediately from the normal ordered form
\beq
U(f) = e^{i\phi^{(-)}(f)} e^{i\phi^{(+)}(f)}\, \,e^{-\fourth (f,f)} \; ,
\eeq
which is a consequence of the  commutation relations (\ref{CCPM})
and the Baker-Campbell-Hausdorff formula \cite{Bourbaki}.

The vacuum functional determines the Hilbert space representation
of the operator algebra. The existence of nontrivial Hilbert
representations depends on invariant functionals of the
test functions $f(\bp)$ which depend only on the neighborhood of $p^+=0$.
Examples  are functionals of
\beq
f_0(p_\perp):=\lim_{p^+\to 0}{f(\bp)\over ip^+}\; ,
\label{FZ}
\eeq
for instance $f_0(0)$ and
\beq
\int d^2p_\perp \vert f_0(p_\perp)\vert^2 \eta(\vert p_\perp\vert^2)\; ,
\eeq
where $\eta(\vert p_\perp\vert^2)$ is some fixed positive function.

A dense set of states  can be represented
by sets of functions $\psi^{(n)}(\bx_1,\dots \bx_n)$  in the $n$-fold tensor
product of $\hat \S(\Re^3)$, because the states
\beq
\Psi := \sum_n \int d^3\bx_1\cdots \int d^3\bx_n
\phi(\bx_1)\dots \phi(\bx_n)\Omega \psi^{(n)}(\bx_1,\dots \bx_n)\; ,
\eeq
are dense in $\H$.
The scalar product $\<\Psi_2\vert \Psi_1\>$  of the states $\Psi_2$
and $\Psi_1$ is therefore related to the vacuum distribution
\beq
W_n(\bx_1,\dots,\bx_n):=E[\phi(\bx_1)\cdots\phi(\bx_n)]
\eeq
by
\beqarray
&&\<\Psi_2\vert\Psi_1\>= \sum_{n,m}\int d^3\bx_1\cdots \int d^3\bx_{n+m}\cr\cr
&&\times\bar \psi_2^{(n)}(\bx_1,\dots,\bx_n)W_{n+m}(\bx_1,\dots, \bx_{n+m})
\psi_1^{m}(\bx_{n+1},\dots, \bx_{n+m})\; .
\label{WF}
\eeqarray
The right-hand side of eq.~(\ref{WF}) is a positive invariant inner product
of functions in $\oplus_n \hat \S^{\otimes n}$.
 We have by completion
a realization of the Hilbert space  by direct sums of multi-point functions.
The inner product of the functions representing states is determined by the
vacuum distributions.

Nontrivial vacuum structures become apparent when we consider  the
vacuum expectation values of the normal ordered operator
\newpage
\beqarray
&&:U(f): = e^{i\phi^{(-)}(f)} e^{i\phi^{(+)}(f)}= U(f)e^{\fourth(f,f)}\cr\cr
&&=\sum_n{i^n\over n!}\int d^3\bx_1\cdots\int d^3\bx_n
:\phi(\bx_1)\cdots\phi(\bx_n):f(\bx_1)
\cdots f(\bx_n) \; .
\eeqarray

The cluster structure of the vacuum requires that
the normal ordered  vacuum distributions
\beq
w_n(\bx_1,\dots,\bx_n):=E[:\phi(\bx_1)\cdots\phi(\bx_n):]
\label{NWF}
\eeq
are products of ``truncated vacuum distributions''\cite{Haag.CC},
$w_T^{(n)}(\bx_1,\dots,\bx_n)$, which vanish in the limit of large spacelike
separations of any two points. They can be defined  by
\beq
E[:U(f):]=\exp \left[ \sum_n i^n w_T^{(n)}(f)\right]\; ,
\label{TVAC}
\eeq
where
\beq
w_T^{(n)}(f):= {1\over n!}
\int d^3\bx_1\cdots\int d^3\bx_n w_T^{(n)}(\bx_1,\dots,\bx_n)f(\bx_1)
\cdots f(\bx_n)\; .
\label{TWF}
\eeq
The distributions (\ref{NWF}) can be obtained from (\ref{TVAC}) by
functional differentiation.

Nonvanishing  distributions $w_T^{(n)}(f)$ represent  nontrivial  vacuum
structure. The requirement (\ref{PPL}) that  the operator $[P^+,\phi(f)]$
annihilates the vacuum in all representations is satisfied when the vacuum
distributions $w_T^{(n)}(f)$  involve only the restriction  (\ref{FZ}).
\beqarray
w_T^{(n)}(f)&:=& {1\over n!}
\int d^3\bp_1\cdots\int d^3\bp_n \hat w_T^{(n)}(p_{1\perp},\dots,p_{n\perp})
\delta(p_{1\perp}+\dots+p_{n\perp})\cr\cr
&\times &{\delta(p_1^+)\over ip^+}f(\bp_1)
\cdots {\delta(p_1^+)\over ip^+}f(\bp_n)\; .
\eeqarray
These vacuum structures are formally the same as those of a
nonrelativistic Bose gase in two space dimensions.\cite{Araki.W}
The positivity of the spectrum of the Hamiltonian $P^-$  together with the
condition that
the Hamiltonian must annihilate the vacuum led to the
requirement (\ref{PPM}), $[P^-,\phi^{(+)}(f)]\Omega=0$. This requirement
relates the dynamics specified by the Hamiltonian to a particular vacuum
structure.

Simple examples of a nontrivial vacuum are obtained when only
$w_T^{(1)}(f)$ and/or $w_T^{(2)}(f)$ are different from zero, for instance,
\beq
w_T^{(1)}(f)=\varphi_0\,f_0(0)\; ,
\eeq
for some real parameter $\varphi_0$, and
\beq
w_T^{(2)}(f)= {1 \over 2}\int d^2p_\perp \vert f_0(p_\perp)\vert^2
\eta(\vert p_\perp\vert^2)\; .
\eeq
for some fixed function $\eta$. Such simple Gaussian vacua may be the basis of
non-Fock perturbative approximations.\\

\vspace{5mm}
\sect{Summary and Conclusions}
\setcounter{equation}{0}
\normalsize

A superficial inspection of the Fock-space formalism indicates that the
Fock vacuum is not a stationary state  when there are pure creation terms in
the
normal ordered interaction Hamiltonian.
In an equal-time Hamiltonian formalism such terms are always present
when the Hamiltonians are derived from  local Lagrangeans.
Since the Fock vacuum is
the only translationally invariant state vector in the Fock space, the physical
vacuum is the cyclic vector of an inequivalent Hilbert-space representation
of the field algebra.

In light-front Hamiltonian dynamics it appears superficially that pure creation
terms
in the Hamiltonian cannot exist
and that the Fock vacuum is therefore the physical vacuum.
This observation rests on the requirement that the sum $p_1^+ + \dots +p_n^+$
vanishes which implies  that  each individual $p_i^+$ vanishes.
For a  more careful analysis it is essential
to note that the fields are  defined as distributions over Schlieder-Seiler
test functions and that $\delta(p^+)/p^+$ is well defined as a distribution.
It follows that nontrivial vacuum structures  exist also in Hamiltonian
light-front dynamics. When a light-front Hamiltonian that is derived formally
from a Lagrangean always possible to suppress the pure creation terms without
affecting the null-plane invariance. It remains to be seen what if any role the
vacuum structures play in approximations to \Poincare invariant field theories.
The role of quark condensates may give an indication.
They can be accommodated in this framework, but only in the presence of a
nontrivial vacuum.

\vspace{5mm}
\sect{Acknowledgments}
\normalsize

\doeack

\end{document}